# Cations determine the mechanism and selectivity of alkaline ORR on Pt(111)

Tomoaki Kumeda[1], Laura Laverdure[2], Karoliina Honkala[2], Marko M. Melander[2,*], and Ken Sakaushi[1,*]

[1]Center for Green Research on Energy and Environmental Materials, National Institute for Materials Science, 1-1 Namiki, Tsukuba, Ibaraki 305-0044, Japan

[2]Department of Chemistry, Nanoscience Center, University of Jyväskylä, P.O. Box 35, FI-40014 Jyväskylä, Finland

*To whom correspondence should be addressed.

E-mail: sakaushi.ken@nims.go.jp (K.S.), marko.m.melander@jyu.fi (M.M.M.)

**Abstract:** The proton-coupled electron transfer (PCET) mechanism of the oxygen reduction reaction (ORR) is a long-standing enigma in electrocatalysis. Despite decades of research, the factors determining the microscopic mechanism of ORR-PCET as a function of pH, electrolyte, and electrode potential remain unresolved, even on the prototypical Pt(111) surface. Herein, we integrate advanced experiments, simulations, and theory to uncover the mechanism of the cation effects on alkaline ORR on well-defined Pt(111). We unveil a dual-cation effect where cations simultaneously determine i) the active electrode surface by controlling the formation of Pt–O and Pt–OH overlayers and ii) the competition between inner- and outer-sphere PCET steps. The cation-dependent transition from Pt–O to Pt–OH determines the ORR mechanism, activity, and selectivity. These findings provide direct evidence that the electrolyte affects the ORR mechanism and performance, with important consequences for the practical design of electrochemical systems and computational catalyst screening studies. Our work highlights the importance of complementary insight from experiments and simulations to understand how different components of the electrochemical interface contribute to electrocatalytic processes.

The oxygen reduction reaction (ORR) is a pivotal reaction in green electrochemical technologies and one of the most important model electrocatalytic reactions. While numerous studies aiming to understand the ORR mechanism have been published,[1] the factors determining the microscopic ORR mechanism and performance remain partly unclear. This incomplete understanding hampers the rational development of more active and selective ORR catalysts.

Compared to the ORR in acidic media, the mechanism, selectivity, and activity of the ORR in alkaline media are more sensitive to reaction conditions such as the electrode potential[2] and electrolytes.[3] In general, the ORR on platinum is widely accepted to proceed via four-electron ($4e^-$) inner-sphere (IS) multi-electron/multi-proton transfer (ET/PT) or proton-coupled electron transfer (PCET) steps leading to the overall conversion of $O_2$ to $H_2O$.[2a, 4] The IS pathway is initiated by $O_2$ adsorption, followed by the production of various surface-bound intermediates, such as $O_2^-{}_{ad}$, $HO_{2,ad}$, $OH_{ad}$, and $O_{ad}$, through successive ET/PT or PCET steps, as shown in Figure 1a (top).[5] The IS pathway is prone to inhibition by high coverage of adsorbed oxide species such as $OH_{ad}$ and $O_{ad}$.[6] Several studies have also suggested alternative pathways where alkaline ORR proceeds through IS and outer-sphere (OS) mechanisms in parallel.[4b, 6a, 7] While the IS route begins by adsorption, the OS pathway is initiated by a long-range OS-ET from the surface to $O_{2,sol}$ forming $O_2^-{}_{sol}$, as illustrated in Figure 1a (bottom).[4b, 8] Crucially, the OS pathway can generate peroxide species through a two-electron ($2e^-$) mechanism.[4b, 7] Understanding and controlling this IS/OS competition is paramount for designing selective $4e^-/2e^-$ ORR catalysts.

Generally, the $4e^-/2e^-$ ORR selectivity depends on the electrode material and the electrolyte. The electrode's effect on selectivity is typically explained by $OH_{ad}$ adsorption energies, as discussed in Ref. [1f] and references therein. $OH_{ad}$ has multiple roles as an ORR intermediate, an initial component in the oxide formation, and an inhibitor of $O_2$ adsorption blocking the IS-PCET pathway. OS-ET reactions may also be sensitive to the electrode material due to surface-specific (non-)covalent interactions.[9] Under alkaline conditions, the role of $OH_{ad}$ depends on non-adsorbing species, such as electrolyte cations at the outer Helmholtz plane (OHP).[3a] For instance, $Li^+$ stabilizes OHad by 0.1 eV whereas $K^+$ destabilizes it by ~0.1 eV.[10] A negative correlation between OHad coverage and acidic ORR activity has been identified, suggesting that OHad blocks $O_2$ adsorption and thus IS-ORR.[11] It was recently proposed that cations impact the reversibility and kinetics of the $OH_{ad} \leftrightarrow O_{ad}$ step, thereby controlling the IS-PCET activity.[12] However, connections between OHad coverage, ORR activity and selectivity, and electrolytes in alkaline media remain to be established.[3d, 13]

While the traditional view is that cations only affect OHad stability and IS-PCET pathways, recent studies on alkaline ORR and $CO_2$ reduction reaction (CO2RR) have shown that cations can induce a more dramatic effect, changing the pathway from an IS-PCET to an OS-ET.[14] $OH_{ad}$ both changes the double layer structure and promotes the OS-ORR pathway in alkaline media,[4b] but the role of cations in the OS mechanism has not been fully understood. Overall, cations appear to have multiple roles in alkaline ORR, suggesting that they may impact the IS and OS pathways through the stabilization of OHad and the formation of solvated species such as $O_2^-{}_{sol}$, respectively. However, this suggestion has not been hitherto confirmed.[3d, 13]

Experimental evidence for the IS/OS in alkaline ORR has been obtained from rotating ring disk electrode (RRDE) studies by detecting OS products, $H_2O_2$ in acidic media and $HO_2^-$ in alkaline media, at different potentials.[4b, 6a, 15] RRDE experiments show that the selectivity depends on the pH, as the amount of $O_2^-{}_{sol}$ formed is pH-sensitive. Increasing the pH from 0 to 7 decreases the overpotential for the first OS-ET step ($O_{2,sol}/O_2^-{}_{sol}$) from 1.53 to 0.7 V,[4b, 16] indicating that OS-ET becomes more facile at higher pH. Support for the existence and contribution of the OS pathway has recently been obtained by combining non-steady state voltammetry and scan direction studies (Figure 1b).[17] Positive-going scan (PGS) RRDE experiments, conventionally used to study ORR performance, show that alkaline ORR on Pt proceeds through an IS-4e$^-$ mechanism.[4b, 6a, 18] As surface oxidation to OHad and Oad species proceeds in parallel with IS-ORR on Pt electrodes, the Pt surface is covered by inhibiting oxides at the beginning of negative-going scans (NGSs) around the ORR onset potential. Furthermore, in alkaline conditions, $O_2^-{}_{sol}$, an intermediate of the OS pathway, is formed through OS-ET around the onset potential.[4b, 17] Therefore, NGSs contain direct information on both the IS and OS pathways while PGSs map the IS pathway, as shown previously.[4b, 6a, 18] The cations and potential scan direction have also been shown to change the ORR and CO2RR pathways from an IS-PCET to an OS-ET.[14] Yet, complementary analyses such as highly accurate electrochemical kinetic isotope effect (haEC-KIE) measurements and advanced electrochemical theory and simulations are needed to distinguish ET and PCET steps.[2b]

While experiments have identified several cation effects and proposed the role of OS pathways in alkaline ORR, most computational and theoretical works have focused on IS-PCET pathways and the electronic properties of the electrode material. The understanding of the OS mechanism and cation effects therefore remains limited. Furthermore, explicit solvation, electrode potential, and electrolyte effects have been largely omitted, even though experimental evidence indicates that the entire electrochemical interface contributes to the alkaline ORR and IS-PCET.[1f] Besides the IS mechanism, pioneering studies by Schmickler's group demonstrated that ORR is likely initialized with an OS-ET forming $O_2^-{}_{sol}$ on Au and Ag.[7] However, these results rely on the accuracy of model Hamiltonian and classical molecular dynamics (MD) parametrizations,[7] and first-principles studies are needed to validate the predictions. A "cross-sphere" mechanism for $H_2O_{sol}$ and $OH_{ad}$ reduction to $OH^-_{sol}$ and $H_2O_{ad}$ species on Pt(111) was recently discovered with DFT-MD and enhanced sampling techniques.[14a] These studies challenge the premise of most computational studies by providing evidence that not only IS steps contribute to alkaline ORR on Pt. They do not fully confirm nor reject the role of initial OS-ET or OS-PCET steps as there are no DFT-level studies on the kinetics of OS pathways and how these are affected by cations.

To understand how cations control the IS/OS and PCET/ET selectivity, we provide a joint experimental–computational–theoretical study on cation effects in alkaline ORR on a Pt(111) single-crystal electrode. The voltammetric and haEC-KIE experiments are integrated with grand canonical ensemble (GCE)-DFT simulations, semi-classical harmonic transition state theory (SC-hTST) of nuclear tunneling under electrochemical conditions, and Marcus theory parametrized with constrained DFT-MD (cDFT-MD). By integrating experiments, theory, and simulations to address the same quantities (coverages, KIEs, potential-dependent kinetics, and thermodynamics) on a well-defined surface, we uncover a novel dual-cation effect where the electrolyte simultaneously impacts both IS- and OS-PCET pathways, activity, and selectivity towards $OH^-$ or $OOH^-$ production. The cations control $OH_{ad}$ stability, which in turn controls the ORR mechanism and selectivity. We demonstrate that the electrolyte affects the alkaline ORR, with important implications for fundamental electrocatalysis, the practical design of electrochemical systems, and catalyst screening studies.

## Results and Discussion

*Surface Coverage*

We studied the $OH_{ad}$ coverage on Pt(111) as a function of the electrode potential using cyclic voltammetry (CVs) in $Li^+$- and $K^+$-containing solutions in an Ar atmosphere. The results are shown in Figures 2a and 2b, respectively. The CVs contain four different regions: hydrogen desorption/adsorption between 0.05 and 0.4 V; the double layer charging/discharging between 0.4 and 0.6 V; OH adsorption/desorption between 0.6 and 0.9 V and irreversible oxide formation above 0.9 V. In agreement with previous studies,[3a, 3b, 19] the OH and oxide regions are sensitive to electrolyte cations. In KOH, the Pt–$OH_{ad}$ coverage reaches 0.5 ML at ~0.78 V, followed by $OH_{ad}$ oxidation to $O_{ad}$ above 0.9 V. In LiOH, the $OH_{ad}$ coverage is 0.5 ML at ~0.74 V, and the Pt–$O_{ad}$ formation is shifted to higher potentials than in KOH. Although OH adsorption starts at the same potential (~0.6 V) in both electrolytes, the $OH_{ad}$ coverage in LiOH increases more rapidly at the potential range between 0.6 and 0.8 V compared to KOH, as shown in Figure 2b. This indicates that $OH_{ad}$ is more resistant to oxidation in LiOH than in KOH, in agreement with previous experiments.[3a, 12]

To determine the cations' effect on the coverage and stability of $OH_{ad}$ and $O_{ad}$, we computed the $OH_{ad}$ and $O_{ad}$ adsorption energies in pure water, constructed surface phase diagrams as a function of pH and electrode potential, and estimated the cation effects as differences between the simulated and measured voltammograms (see Supporting Information for details). The simulated coverages without cations qualitatively agree with those measured in KOH during the PGS. In LiOH, the peak corresponding to $O_{ad}$ formation is shifted to higher potentials outside the potential range studied (see Supporting Information Figure S4),[3a] indicating that $Li^+$ either weakens $O_{ad}$ or stabilizes $OH_{ad}$, or both. To quantify the impact of cation-induced (de)stabilization on the adsorbed oxide species, we investigated how the simulated coverages and voltammograms change when the OH and O adsorption energies are manually modified from −0.2 to 0.2 eV. When the OH adsorption was stabilized, the onset of the $OH_{ad}$ peak (~0.6 V) was observed to shift to lower potentials, in contrast to our experiments. Destabilizing $O_{ad}$ relative to $OH_{ad}$ by 0.05−0.1 eV shifts the $O_{ad}$ peak from around 1.04 V to potentials outside the experimental potential range, as observed in the LiOH CV (see Supporting Information Figure S15). With a 0.1 eV $O_{ad}$ destabilization, the calculations predict an OH coverage of 8/9 ML at ~0.95 V, which corresponds well to the measured coverage of ~0.9 ML at 1.05 V in LiOH.

The best fit to the experimental Li$^+$ data is obtained by weakening the Oad energy by 0.05−0.1 eV, which we interpret as a relative stabilization of $OH_{ad}$ compared to $O_{ad}$, as suggested recently by Luo and Koper.[12] The magnitude of this Li$^+$ (de)stabilization agrees well with previous experiments[3a] and DFT-MD simulations.[10] Overall, our results show that even a modest 0.05−0.1 eV $OH_{ad}$ stabilization or $O_{ad}$ destabilization by electrolyte cations substantially impacts the O and OH coverage of the Pt(111) surface near the ORR thermodynamic potential. We also note that the experimental ORR onset potentials in KOH and LiOH (~1.0 V) coincide with the threshold potentials of an O coverage of 5/9 ML and an OH coverage of 8/9 ML (Figure S12). This indicates that ORR cannot proceed when the coverages of PtOH and PtO are too high.

*ORR Kinetics and Mechanism: Experiments*

The adsorbate stability directly impacts the Pt(111) surface state and hence the ORR performance. Figure 3a displays a ~130 mV shift in the half-wave potential between KOH and LiOH voltammograms during the PGS, in agreement with previous studies.[3a] The ORR kinetic region is limited by the OHad region, indicating that the cation-dependent formation and stability of OHad directly affect ORR activity. As shown in Figure 3a, the ORR current exhibits a hysteresis (negative shift of the half-wave potential) between the PGS and NGS. Specifically, the ORR current in KOH is smaller during the NGS than during the PGS; less shift is seen in LiOH. For comparison, we also conducted similar CV and ORR polarization measurements in NaOH and HClO$_4$ solutions, which exhibit a hysteresis comparable to that in KOH and LiOH solutions, respectively (see Supporting Information).

Based on the measured and simulated CVs and coverages, we attribute the hysteresis in KOH and NaOH to the irreversible O$_{ad}$ formation above 0.9 V during the PGS and reduction below 0.9 V during the NGS, as shown in the CVs in Figure 2a and Figure S1. During PGSs, the IS pathway proceeds on the surface, which gradually becomes covered by reversible OH$_{ad}$ (0.6−0.9 V) followed by irreversible O$_{ad}$ (>0.9 V). As the IS pathway is inhibited by high surface coverages,[6a] the lower ORR activity in LiOH compared to KOH during the PGS is attributed to the higher coverage of poisoning OH$_{ad}$.[3a] During the NGS from 1.05 to 0.8 V, the irreversible O$_{ad}$ inhibits the IS pathway. The difference in the surface oxidation state between PGSs and NGSs causes the ORR hysteresis in KOH. In LiOH, since the irreversible O$_{ad}$ is formed at higher potentials (> 1.0 V), less change in the surface oxidation state occurs between PGSs and NGSs in the measured potential range (0.05−1.05 V). When the polarization changes at 1.05 V, O$_2^-{}_{sol}$, the initial intermediate for the OS-ORR pathway, is formed through OS-ET, as schematically shown in Figure 1b.[4b] As discussed below in ORR kinetics and mechanism: Simulations and suggested in Ref. [4b], the OS-ET is promoted by OH$_{ad}$ stabilized in LiOH. Therefore, the NGS reflects the effects of the irreversible O$_{ad}$ in KOH and O$_2^-{}_{sol}$ in LiOH. As the potential scan direction can alter the ORR mechanism from the IS to OS pathway,[4b, 6a, 18] we interpret the smaller ORR hysteresis between the PGS and NGS in LiOH compared to KOH as an indication of the OS pathway being promoted (see Supporting Information pages 11−12 for details).

As oxide formation on Pt depends on the electrolyte and potential range applied, we studied the ORR hysteresis in alkaline solutions: LiOH, NaOH, KOH, and tetramethylammonium hydroxide (TMAOH), with different upper potential limits from 0.9 to 1.2 V. The hysteresis is quantified by shifts in the ORR half-wave potential from PGSs to NGSs shown in Figure 3b (see Supporting Information Figure S4 for Ar-saturated CVs and ORR voltammograms). The ORR hysteresis becomes more pronounced as the upper potential limit increases in all investigated electrolyte conditions. This result indicates that the irreversible Oad species formed at higher potentials block the active sites for the IS pathway during the NGS. The potential shifts in LiOH are smaller than in NaOH, KOH, and TMAOH, regardless of the upper potential limit. Crucially, the hysteresis is present in NaOH, KOH, and TMAOH with upper potential limits of 0.9 and 0.95 V, even though OHad is the major oxide species. Therefore, the irreversible Oad formation depending on the upper potential limit cannot fully explain the ORR hysteresis; electrolyte cations influence the ORR kinetics and IS/OS selectivity.

The formation of peroxide species and the possibility of OS pathways were studied by RRDE experiments as a function of pH (see Supporting Information for details). The RRDE profiles' pH-dependence shown in Supporting Information Figure S6 is consistent with a previous study,[4b] and proves the formation of OS pathway species. Although the IS/OS selectivity depends on the Pt surface morphology, the trends in the ORR kinetic current between 0.6 and 1.0 V in different electrolytes (LiOH < KOH) and for different scan directions (positive > negative in KOH and negative ≥ positive in LiOH) on poly-Pt correlate well with those observed for Pt(111) (see Supporting Information). Poly-Pt is therefore a suitable Pt(111) proxy to study the OS pathway through RRDE experiments.

Figure 3c shows ring currents of RRDE experiments in LiOH, NaOH, KOH, and TMAOH. In NaOH, KOH, and TMAOH, the lack of ring current during the PGS indicates that the reaction follows a 4e$^-$ IS pathway, as reported previously.[6a] In LiOH, the small ring current peak at 0.83 V during the PGS corresponds to the peroxide formation and is typically assigned to OS-ET for Pt-catalyzed ORR (see Supporting Information for details).[9c, 20] Although the peak is ~0.1 V higher than the equilibrium potential for $H_2O_2$ (E° = +0.695 V), this difference is attributed to the lower local $O_2/H_2O_2$ ($HO_2^-$) ratio, which subsequently changes the equilibrium potential according to the Nernst equation.[21] In the NGS, relatively large ring currents are observed in both solutions. The peak ring current in LiOH is approximately four times larger than in KOH. These results strongly suggest that LiOH promotes the 2e⁻ OS process and that $OH_{ad}$ and solvated species, such as $O_2^-{}_{sol}$, accumulate near the surface and significantly contribute to the OS pathway.

The peak ring current during the NGS increases as the upper potential increases, as shown in Figure 3d. This indicates that the amount of dissolved intermediates from the 2e⁻ OS pathway increases near the electrode surface above the ORR onset potential. The cation-dependent peroxide yield during the NGS follows the sequence $Li^+$ > $Na^+$ > $K^+$ > $TMA^+$, regardless of the upper potential limit. This trend is consistent with the cations' hydration energies ($\Delta H_{hyd}$) ($Li^+$ > $Na^+$ > $K^+$ > $TMA^+$).[3a, 19b] This suggests that the strength of cation−$OH_{ad}$ and cation−$H_2O$ interactions, which determine the $OH_{ad}$ stability and interfacial water structure,[3a] are descriptors for the IS and OS competition.

The difference between OS and IS steps and mechanistic insight into the ET and PCET steps in KOH and LiOH aqueous solutions were extracted from haEC-KIE measurements [1e, 2b, 22] (see Supporting Information for details). The KIE rate constant ratios ($K^{H/D}$) and the transfer coefficients ($α_{H/D}$) at representative potentials extracted from a Tafel analysis of haEC-KIE measurements are given in Supporting Information Table S1. The haEC-KIE analysis shows that the rate-determining step (RDS) depends on the scan direction and the electrolyte. In KOH, during the PGS, the $K^{H/D}$ values at 0.93 $V_{RHE}$ and 0.83 $V_{RHE}$ are 1.5 and 1.6, respectively, which indicates that the RDS either involves a proton transfer process or is influenced by water dynamics.[23] The corresponding $K^{H/D}$ values for the NGSs are 0.8 and 1.2. While electrochemical KIEs below 1 have traditionally been taken as signatures of pure ET, it has been recently shown that, in ORR, KIEs<1 result from the greater stability of $OD_{ad}$ compared to $OH_{ad}$.[24] Hence, at 0.93 V in KOH, OH/OD adsorption determines the KIE, while at lower potentials other factors, presumably kinetic, also contribute to the KIE. The $K^{H/D}$ values in aqueous LiOH for the NGSs and PGSs are between 1.5 and 1.7. in the kinetic region at 0.88 $V_{RHE}$, indicating that the RDS involves a PCET step and that OD/OH stability is not the KIE-controlling factor at these potentials.

*ORR Kinetics and Mechanism: Simulations*

To understand the atomistic underpinnings of the experimental results, we studied both the IS and OS pathways by examining the reaction kinetics, thermodynamics, and KIEs for the $O_2 \rightarrow OOH$ and $O \rightarrow OH$ steps, which are typically regarded as the slowest electrochemical PCET steps in alkaline ORR on Pt(111).[2a, 12, 24b] Another possible KIE- and rate-determining step, OH desorption, was recently considered in detail[24b] and thus we do not consider it here. For the IS-PCET pathway, we computed the EC-KIEs as a function of the electrode potential using GCE-DFT and SC-hTST adapted for constant potential calculations, as summarized in Supporting Information Figure S18a (see also Supporting Information). The EC-KIEs were computed at 1.13 $V_{RHE}$ and 0.73 $V_{RHE}$, both at pH=13, corresponding to the limits of the experimental conditions in this work. At 1.13 V (overpotential −0.1 V), the IS-PCET free energy barriers are small being below 0.3 eV, and they decrease when the overpotential decreases to −0.5 V, as shown in Table 1. The reaction barriers and energies show that the IS-PCET steps proceed readily on Pt(111) under alkaline conditions and are consistent with previous computational studies.[2a, 14a] The barrier for the second PCET step, from $O_{ad}$ to $OH_{ad}$, is twice that of the first step and is expected to be the RDS at small overpotentials, in agreement with the measured Tafel slopes. Under more reducing conditions, the situation is less clear because the free energy barriers for both reactions are almost identical and smaller than 0.1 eV. However, the $O_{2,ad} \rightarrow OOH_{ad}$ step is expected to be slightly faster due to nuclear quantum effects discussed below.

Computed and measured KIEs show better agreement for the $O \rightarrow OH_{ad}$ step than for the $O_2 \rightarrow OOH_{,ad}$ step suggesting that the former is the KIE-determining step for the IS-PCET ORR mechanism on Pt(111) with LiOH and KOH electrolytes. Despite a rather small KIE, nuclear quantum effects can significantly influence reaction rates. The tunneling correction ($\kappa \approx 4$) for $O_2 \rightarrow OOH_{,ad}$ at 0.73 V makes this step slightly faster than $O \rightarrow OH_{ad}$. The tunneling corrections vary due to subtle changes in the reaction mechanism and bond lengths at different potentials. For instance, at 0.73 V, the $O_2 \rightarrow OOH_{ad}$ transition state involves both water reorganization and high vibration frequency proton transfer motions, whereas at 1.13 V, the transition state only involves water reorganization, which is followed by a facile proton transfer with no additional barrier. The analysis of the results (see Supporting Information) shows that the KIEs of the elementary steps studied do not strongly depend on the presence of cations.

The OS-ET pathways were studied by simulating the initial $O_{2,sol} \rightarrow O_2^-{}_{sol}$ OS-ET step using Marcus theory parametrized with cDFT-MD simulations (see Figure S15a for details). Our results show that OS-ET in pure water is very slow and has a high kinetic barrier of ~1.3 eV (see Supporting Information). To explore the impact of cations on the OS pathways, the cDFT-MD simulations were repeated with $Li^+$ or $K^+$ (Figure S15d). Initially, $O_{2,sol}$ and $O_2^-{}_{sol}$ form complexes with both cations, but only the $K^+$–$O_2$ complex remains stable throughout the 10 ps simulation. $Li^+$ is more reactive, and the initial $Li^+$–$O_2$ complex reacts with an interfacial water molecule, leading to the formation of $Li^+$–OOH and $OH_{ad}$ within 1 ps.

The calculations show that $O_2^-$–sol is substantially more reactive than $O_{2,sol}$. In both LiOH and KOH, the $O_2^-{}_{sol}$ coordinates with the corresponding cation ($M^+$) for a few picoseconds before abstracting a proton from a water molecule close to the Pt surface, leading to the formation of $M^+$–$OOH^-$ and $OH_{ad}$. Due to the unstable and highly reactive nature of $O_2^-{}_{sol}$ in cation-containing solutions, we were unable to extract kinetics from these simulations. However, qualitative observations show that both the $Li^+$ and $K^+$ have a clear impact on the OS-ET ORR steps as they form complexes with $O_{2,sol}$, $O_2^-{}_{sol}$, and peroxide species, and control which species form during the simulation. Our simulations show that $Li^+$ activates $O_{2,sol}$ and $O_2^-{}_{sol}$ towards the formation of peroxide species more strongly than $K^+$ or pure water. We interpret this as evidence for $Li^+$ catalyzing the OS pathways in agreement with our experimental measurements. The simulations also show that $O_{2,sol}$/$O_2^-{}_{sol}$ abstract protons exclusively from near-surface water molecules producing $OH_{ad}$, which suggests that peroxide formation cannot take place without interfacial water molecules and vacant Pt sites for OH adsorption.

*Understanding the Electrolyte-Dependency in Alkaline ORR on Pt(111)*

Our experimental and computational results explain why the alkaline ORR mechanism on Pt(111) depends strongly on the electrolyte. We extend the cation-dependent OHad stabilization model suggested by Marković's group,[3a] by showing that cations directly control both the IS and OS pathways. For the IS pathway, our results demonstrate that the OHad/Oad stability depends on the cation. Compared to $K^+$, $Li^+$ destabilizes Oad relative to OHad by ~0.05-0.1 eV in accordance with recent findings.[12] $Li^+$ and $K^+$ were found to promote OS pathways, which are unfavourable in pure water. The presence of $Li^+$ makes $O_{2,sol}$ more likely to participate in OS-ET and OS-PCET, while in $K^+$ solutions, the more stable $K^+$–$O_{2,sol}$ does not participate in any reaction during the simulation timescale. Any $O_2^-{}_{sol}$ formed through OS-ET in either pure water or electrolyte solutions readily reacts with a water molecule near the surface, resulting in the formation of $OH_{ad}$ followed by either $HO_2^-{}_{sol}$ or $H_2O_{2,sol}$. Together, these observations explain why the formation of the peroxide species coincides with the adsorption potentials of O and OH in electrolyte solutions. Overall, these findings show that the cations have a dual role in alkaline ORR: they affect both the IS and OS pathways through surface coverages and by controlling OS-ET and OS-PCET steps.

The above mechanisms, summarized in Figure 4, allow us to explain the cation-dependent hysteresis shown in Figure 3 and previous studies.[18] The differences in the polarization curves of Figure 3a are attributed to i) the cation-dependent O/OH coverages (Figures 2 and S15) controlling surface blocking determining the IS-ORR activity, and ii) the OS-ORR activity evidenced by the ring currents in Figure 3c. During the PGS, the ORR proceeds primarily via an IS pathway in both LiOH and KOH solutions, and the cation-dependency arises from the stabilization of $OH_{ad}$ relative to $O_{ad}$, blocking $O_2$ adsorption. The $K^{H/D}$ value of ~1.5 in KOH indicates that, during the PGS, the RDS is water dynamics or a PCET step, most likely O conversion to OH. In contrast, during the NGS, the ORR mechanisms are different in KOH and LiOH solutions. During the NGS in KOH, the ORR kinetic current is significantly lower than during the PGS, and stable OHad or Oad species inhibit the IS-ORR pathway by surface blocking (Figure 4, top). In LiOH, PtO is not formed in the potential range considered because $OH_{ad}$ is more stable. $Li^+$ also facilitates the formation and accumulation of solvated species, such as $O_2^-{}_{sol}$ and $HO_{2,sol}$, near the surface through OS steps during the NGS (Figure 4, bottom). A $K^{H/D}$ of ~1.5 for the NGS in LiOH indicates that the RDS involves water dynamics or a proton transfer. This in turn, suggests that either an OS-PCET or an associative mechanism with both IS and OS steps might create $HO_2^-$ ($HO_2^-{}_{ad}/O_2^-{}_{ad} \leftrightarrow HO_2^-{}_{sol}/O_2^-{}_{sol}$).[17b] The computed KIEs for the $O_{ad} \to OH_{ad}$ step provide are in good agreement with the measured $K^{H/D}$ and strongly support the recent conclusion[12] that this step is not only the RDS of alkaline ORR but that its kinetics control the overall ORR activity.

## Conclusion

We addressed the ORR mechanisms, kinetics, and selectivity in alkaline media on a well-defined system, Pt(111), with different cations using experimental and computational methods. Our results show that electrolyte cations qualitatively and quantitatively modify the alkaline ORR chemistry. Our study uncovers a novel dual-cation effect, where the electrolyte simultaneously impacts both the IS and OS-PCET pathways by 1) controlling the formation of surface-blocking adsorbed species ($OH_{ad}$ and $O_{ad}$) through $OH_{ad}$ stabilization relative to $O_{ad}$ and 2) stabilizing $O_2^-{}_{sol}/HO_2^-{}_{sol}$, which both influence the selectivity between IS and OS pathways. Overall, our findings have important implications for the role of cations and outer-sphere pathways in ORR and electrocatalysis in general, emphasizing the need to account for the full complexity of the electrochemical interface and various mechanistic scenarios when designing electrocatalytic systems.

*Supporting Information*

The authors have cited additional references within the Supporting Information.

*Acknowledgements*

T.K and K.S were supported by Japan Society for the Promotion of Science (JSPS/KAKENHI, 19H05460, 21J00688, and 22KJ3237). M.M.M acknowledges support by the Academy of Finland (projects #338228 and #30785). M.M.M and K.H acknowledge the Janes and Aatos Erkko foundation for support to the LACOR project. M.M.M., K.H., and L.L. acknowledge support by the Academy of Finland to projects #317739. The computational resources are provided by CSC—IT Center for Science, Espoo, Finland (https://www.csc.fi/en/). T.K. and K.S. are indebted to NIMS, Japan.

[1] a) A. Kulkarni, S. Siahrostami, A. Patel, J. K. Norskov, Chem. Rev. 2018, 118, 2302-2312; b) J. Huang, M. Eikerling, Curr. Opin. Electrochem. 2019, 13, 157-165; c) N. Ramaswamy, S. Mukerjee, Chem. Rev. 2019, 119, 11945-11979; d) K. Sakaushi, A. Lyalin, S. Tominaka, T. Taketsugu, K. Uosaki, ACS Nano 2017, 11, 1770-1779; e) K. Sakaushi, M. Eckardt, A. Lyalin, T. Taketsugu, R. J. Behm, K. Uosaki, ACS Catal. 2018, 8, 8162-8176; f) Y. Yang, C. R. Peltier, R. Zeng, R. Schimmenti, Q. Li, X. Huang, Z. Yan, G. Potsi, R. Selhorst, X. Lu, W. Xu, M. Tader, A. V. Soudackov, H. Zhang, M. Krumov, E. Murray, P. Xu, J. Hitt, L. Xu, H. Y. Ko, B. G. Ernst, C. Bundschu, A. Luo, D. Markovich, M. Hu, C. He, H. Wang, J. Fang, R. A. DiStasio, Jr., L. F. Kourkoutis, A. Singer, K. J. T. Noonan, L. Xiao, L. Zhuang, B. S. Pivovar, P. Zelenay, E. Herrero, J. M. Feliu, J. Suntivich, E. P. Giannelis, S. Hammes-Schiffer, T. Arias, M. Mavrikakis, T. E. Mallouk, J. D. Brock, D. A. Muller, F. J. DiSalvo, G. W. Coates, H. D. Abruna, Chem. Rev. 2022, 122, 6117-6321.

[2] a) S. Liu, M. G. White, P. Liu, J. Phys. Chem. C 2016, 120, 15288-15298; b) K. Sakaushi, A. Lyalin, T. Taketsugu, K. Uosaki, Phys. Rev. Lett. 2018, 121, 236001.

[3] a) D. Strmcnik, K. Kodama, D. van der Vliet, J. Greeley, V. R. Stamenkovic, N. M. Markovic, Nat. Chem. 2009, 1, 466-472; b) D. Strmcnik, D. F. van der Vliet, K. C. Chang, V. Komanicky, K. Kodama, H. You, V. R. Stamenkovic, N. M. Marković, J. Phys. Chem. Lett. 2011,

2, 2733-2736; c) J. Suntivich, E. E. Perry, H. A. Gasteiger, Y. Shao-Horn, Electrocatalysis 2012, 4, 49-55; d) B. Garlyyev, S. Xue, M. D. Pohl, D. Reinisch, A. S. Bandarenka, ACS Omega 2018, 3, 15325-15331; e) T. Kumeda, K. Sakaushi, Curr. Opin. Electrochem. 2022, 36, 101121.

[4]     a) A. J. Appleby, J. Electroanal. Chem. 1993, 357, 117-179; b) N. Ramaswamy, S. Mukerjee, J. Phys. Chem. C 2011, 115, 18015-18026.

[5]     M. T. M. Koper, Chem. Sci. 2013, 4, 2710-2723.

[6]     a) N. M. Marković, H. A. Gasteiger, P. N. Ross, J. Phys. Chem. 1996, 100, 6715-6721; b) T. J. Schmidt, V. Stamenkovic, J. P. N. Ross, N. M. Markovic, Phys. Chem. Chem. Phys. 2003, 5, 400-406.

[7]     A. Ignaczak, E. Santos, W. Schmickler, Curr. Opin. Electrochem. 2019, 14, 180-185.

[8]     a) J. O. Bockris, A. J. Appleby, in Assessment of Research Needs for Advanced Fuel Cells (Ed.: S. S. Penner), Pergamon Press, Oxford, 1986, pp. 95-135; b) A. J. Appleby, in Comprehensive Treatise of Electrochemistry, Vol. 7 (Eds.: B. E. Conway, J. O. M. Bockris, E. Yeager, S. U. M. Khan, R. E. White), Plenum Press, New York, 1983, pp. 173-239.

[9]     a) E. Santos, R. Nazmutdinov, W. Schmickler, Curr. Opin. Electrochem. 2020, 19, 106-112; b) A. Ignaczak, R. Nazmutdinov, A. Goduljan, L. M. de Campos Pinto, F. Juarez, P. Quaino, G. Belletti, E. Santos, W. Schmickler, Electrocatalysis 2017, 8, 554-564; c) N. Ramaswamy, S. Mukerjee, Adv. Phys. Chem. 2012, 2012, 491604; d) D. Q. Liu, M. Kang, D. Perry, C. H. Chen, G. West, X. Xia, S. Chaudhuri, Z. P. L. Laker, N. R. Wilson, G. N. Meloni, M. M. Melander, R. J. Maurer, P. R. Unwin, Nat. Commun. 2021, 12, 7110.

[10]    H. H. Kristoffersen, K. Chan, T. Vegge, H. A. Hansen, Chem. Commun. 2020, 56, 427-430.

[11] a) M. B. Vukmirovic, J. Zhang, K. Sasaki, A. U. Nilekar, F. Uribe, M. Mavrikakis, R. R. Adzic, Electrochim. Acta 2007, 52, 2257-2263; b) V. R. Stamenkovic, B. Fowler, B. S. Mun, G. Wang, P. N. Ross, C. A. Lucas, N. M. Markovic, Science 2007, 315, 493-497; c) Y.-J. Deng, M. Arenz, G. K. H. Wiberg, Electrochem. Commun. 2015, 53, 41-44; d) H. Tanaka, S. Sugawara, K. Shinohara, T. Ueno, S. Suzuki, N. Hoshi, M. Nakamura, Electrocatalysis 2014, 6, 295-299; e) T. Ueno, H. Tanaka, S. Sugawara, K. Shinohara, A. Ohma, N. Hoshi, M. Nakamura, J. Electroanal. Chem. 2017, 800, 162-166; f) T. Kumeda, H. Tajiri, O. Sakata, N. Hoshi, M. Nakamura, Nat. Commun. 2018, 9, 4378.

[12] M. Luo, M. T. M. Koper, Nat. Catal. 2022, 5, 615-623.

[13] R. Rizo, E. Herrero, J. M. Feliu, Phys. Chem. Chem. Phys. 2013, 15, 15416-15425.

[14] a) Y. Li, Z. F. Liu, J. Phys. Chem. Lett. 2021, 12, 6448-6456; b) V. Sinha, E. Khramenkova, E. A. Pidko, Chem. Sci. 2022, 13, 3803-3808.

[15] B. B. Blizanac, P. N. Ross, N. M. Markovic, J. Phys. Chem. B 2006, 110, 4735-4741.

[16] B. B. Blizanac, P. N. Ross, N. M. Markovic, Electrochim. Acta 2007, 52, 2264-2271.

[17] a) A. M. Gómez-Marín, J. M. Feliu, ChemSusChem 2013, 6, 1091-1100; b) A. Gómez-Marín, J. Feliu, T. Edson, ACS Catal. 2018, 8, 7931-7943.

[18] Q. Wang, H. Guesmi, S. Tingry, D. Cornu, Y. Holade, S. D. Minteer, ACS Energy Lett. 2022, 7, 952-957.

[19] a) M. Nakamura, Y. Nakajima, N. Hoshi, H. Tajiri, O. Sakata, ChemPhysChem 2013, 14, 2426-2431; b) T. Kumeda, R. Kubo, N. Hoshi, M. Nakamura, ACS Appl. Energy Mater. 2019, 2, 3904-3909.

[20] I. Katsounaros, W. B. Schneider, J. C. Meier, U. Benedikt, P. U. Biedermann, A. A. Auer, K. J. Mayrhofer, Phys Chem Chem Phys 2012, 14, 7384-7391.

[21]     A. Ignaczak, R. Nazmutdinov, A. Goduljan, L. Moreira de Campos Pinto, F. Juarez, P. Quaino, E. Santos, W. Schmickler, Nano Energy 2016, 26, 558-564.

[22]     a) K. Sakaushi, Faraday Discuss. 2019, 221, 428-448; b) K. Sakaushi, Phys. Chem. Chem. Phys. 2020, 22, 11219-11243.

[23]     L. Rebollar, S. Intikhab, J. D. Snyder, M. H. Tang, J. Phys. Chem. Lett. 2020, 11, 2308-2313.

[24]     a) C. Hartnig, M. T. M. Koper, J. Electroanal. Chem. 2002, 532, 165-170; b) Y. Yang, R. G. Agarwal, P. Hutchison, R. Rizo, A. V. Soudackov, X. Lu, E. Herrero, J. M. Feliu, S. Hammes-Schiffer, J. M. Mayer, H. D. Abruna, Nat. Chem. 2023, 15, 271-277.

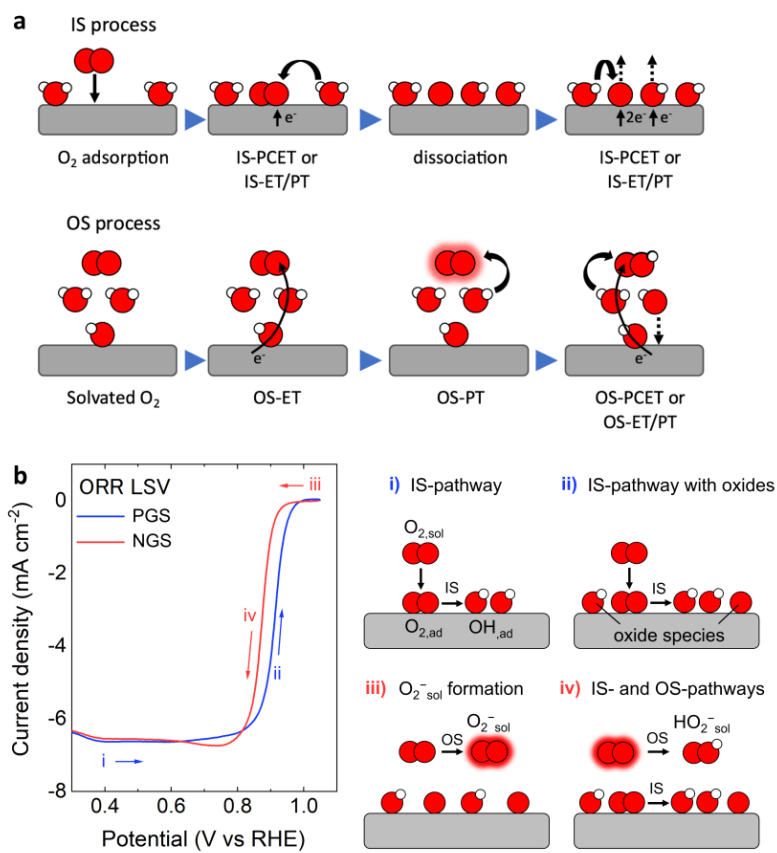

**Figure 1.** a) ORR mechanisms in alkaline media. (top) IS-ORR via the associative pathway is initiated by the adsorption of $O_2$, which then forms $HO_{2,ad}$ through an IS-PCET or IS-ET/IS-PT step. $HO_{2,ad}$ dissociates into $OH_{ad}$ and $O_{ad}$. Finally, $OH^-_{ad}$ is formed through an IS-PCET or IS-ET/IS-PT, then $OH^-_{ad}$ desorbs, forming $OH^-_{sol}$. (bottom) OS-ORR is initiated by a long-range OS-ET from the electrode to $O_{2,sol}$ via $OH_{ad}$, yielding $O_2^-{}_{sol}$ (highlighted $O_2$ molecule). The very reactive $O_2^-{}_{sol}$ abstracts a proton from a nearby water molecule and receives an electron through an OS-PCET or OS-PT/OS-ET resulting in $HO_2^-{}_{sol}$ and either $OH^-_{sol}$ or $OH_{ad}$. b) ORR linear sweep voltammogram (LSV) for Pt(111) with rotating disk electrode technique and the corresponding schematic explanation (i-iv) of the LSV of Pt in alkaline solution during positive-going scans (PGSs) and negative-going scans (NGSs). (i) Before the surface oxidation during the PGS, the inner-sphere (IS) $4e^-$ pathway proceeds. (ii) The IS pathway proceeds after the surface is oxidized and covered by adsorbed oxide species during the PGS. (iii) Around the onset potential (~1 V vs. RHE), the surface is covered by $OH_{ad}$ and $O_{ad}$. $OH_{ad}$ promotes the outer-sphere (OS) electron transfer, forming $O_2^-{}_{sol}$. (iv) During the NGS, the surface is covered by $OH_{ad}$ and $O_{ad}$, and the IS- and OS-ORR pathways proceed in parallel.

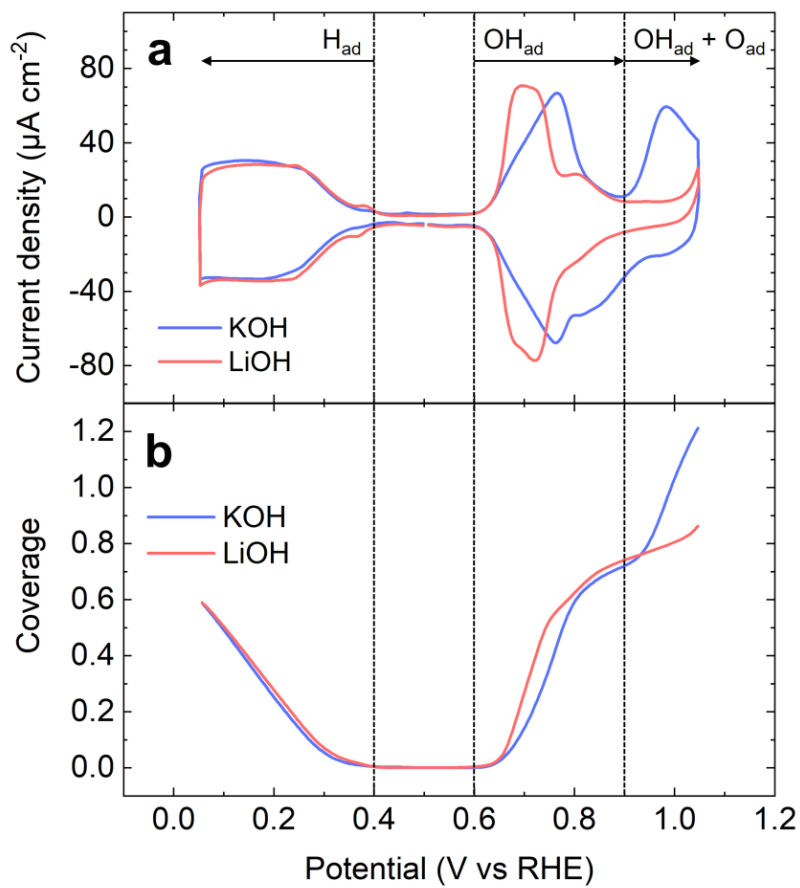

**Figure 2.** a) Ar-saturated CVs of Pt(111) in 0.1 M KOH and 0.1 M LiOH in $H_2O$ with an upper potential limit at 1.05 V (RHE). b) The corresponding $H_{ad}$, $OH_{ad}$, and $O_{ad}$ coverages extracted from the PGS. The scan speed was 50 mV s$^{-1}$.

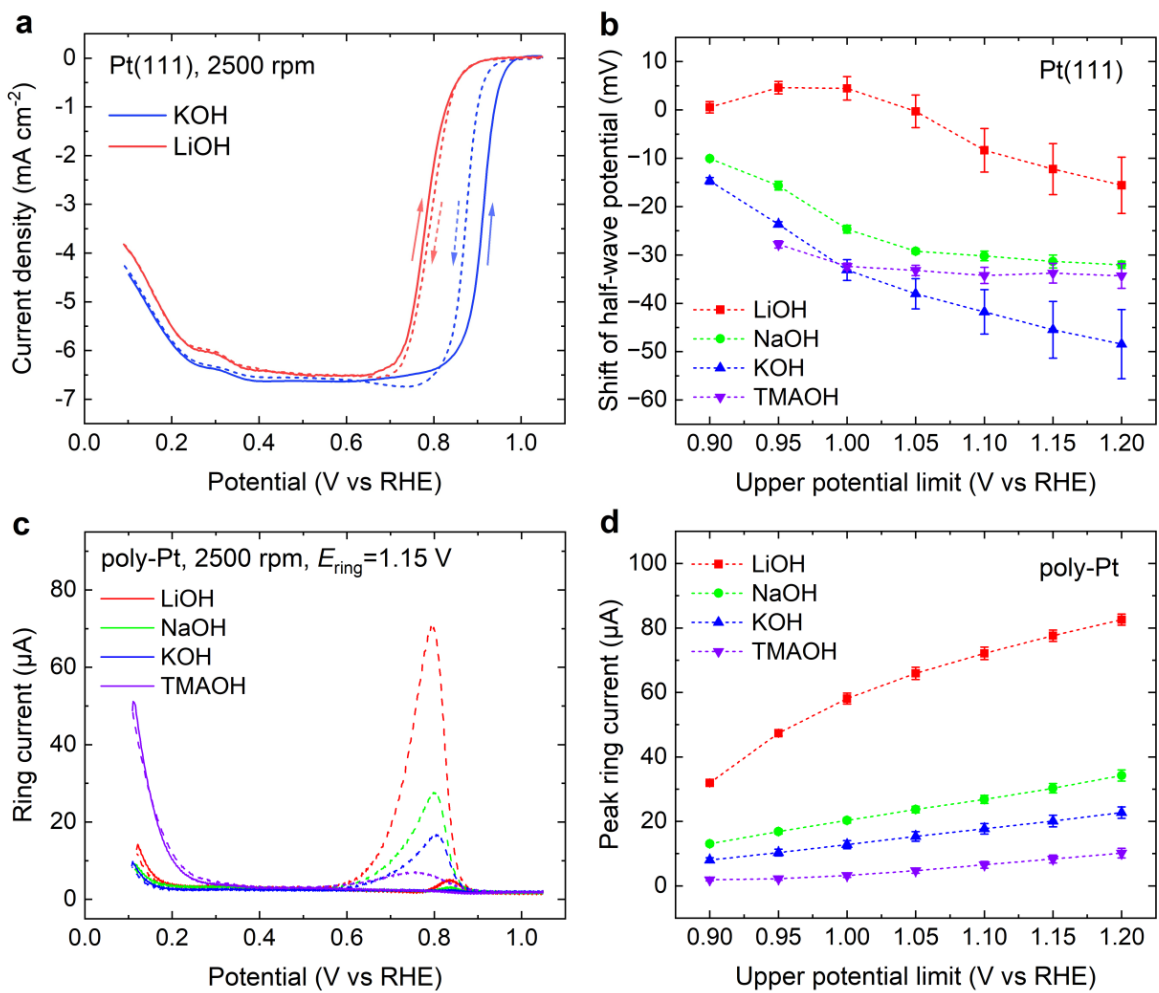

Figure 3. a) $O_2$-saturated voltammograms of Pt(111) during the PGS (solid line) and NGS (dashed line) of Pt(111) in aqueous solutions of KOH and LiOH with an upper potential limit at 1.05 V (RHE). b) Shifts of the ORR half-wave potential from PGS to NGS of Pt(111) in LiOH, NaOH, KOH, and TMAOH with different upper potential limits. c) Ring current from $O_2$-saturated voltammograms of poly-Pt ring and disk electrodes in LiOH, NaOH, KOH, and TMAOH during PGSs (solid line) and NGSs (dashed line) with an upper potential limit at 1.05 V (RHE). The constant potential applied to the ring electrode is 1.15 V. d) Peak ring current in $O_2$-saturated LiOH, NaOH, KOH, and TMAOH during the NGS with different upper potential limits. In (a) and (c), the working electrode rotation rate is 2500 rpm, and the potential scanning rate is 50 mV s$^{-1}$. All electrolyte concentrations are fixed to 0.1 M.

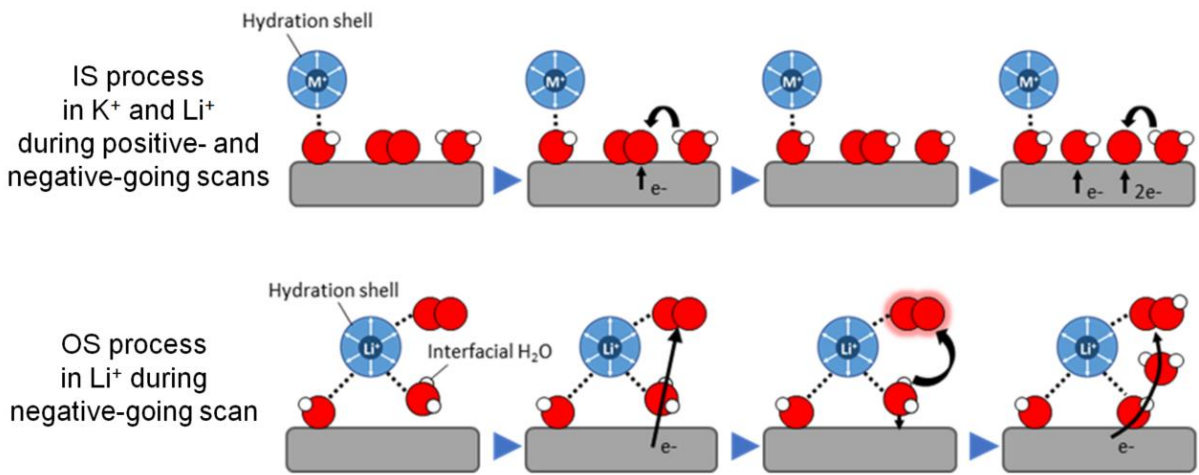

**Figure 4.** Schematic mechanisms for alkaline ORR on Pt involving IS and OS electron transfer and proton-coupled electron transfer. The IS mechanism during PGSs and NGSs in $K^+$- and $Li^+$-containing solutions (top). The OS mechanism during the NGS in $Li^+$-containing solution (bottom). The white arrows around the cations denote the direction of the hydration water dipoles. The dotted lines between the hydrated cations, $OH_{ad}$ ($O_{ad}$), and ORR intermediates denote non-covalent interactions.

**Table 1.** Computed energy barriers ($E^{\ddagger}$), grand free energy barriers ($\Omega^{\ddagger}$), reaction energies ($\Delta E$), and KIEs for IS-PCET steps. KIE TST indicates KIE computed from GCE-DFT energies using the harmonic TST based on vibrational frequencies (Supporting Information). KIE SC-hTST corresponds to the semiclassical harmonic TST expression accounting for the tunneling prefactor $\kappa$. $\omega^{\ddagger}$ is the imaginary vibrational frequency at the TS.

| Reaction | Potential [$V_{RHE}$] | $E^{\ddagger}$ [eV] | $\Omega^{\ddagger}$ [eV] | $\Delta E$ | $\omega^{\ddagger}$ [cm$^{-1}$] | $\kappa$ | KIE TST | KIE SC-hTST |
|---|---|---|---|---|---|---|---|---|
| $O_2 \to OOH_{ad}$ | 0.73 | 0.11 | 0.07 | -0.11 | 1821 | 3.9 | 5.8 | 4.3 |
| | 1.13 | 0.14 | 0.12 | -0.01 | 734 | 1.7 | 3.5 | 4.3 |
| $O \to OH_{ad}$ | 0.73 | 0.12 | 0.06 | -0.56 | 455 | 1.2 | 2.1 | 2.1 |
| | 1.13 | 0.29 | 0.26 | -0.47 | 804 | 2.0 | 1.5 | 1.6 |